\documentclass{aa} 
\usepackage{graphicx}
\newcommand{\teff}{\mbox{${T}_{\rm eff}\,\,$}}

\newcommand{\msol}{\mbox{${\rm M}_{\odot}\,\,$}}

\newcommand{\lil}{\mbox{$^6{\rm Li}\,\,$}}
\newcommand{\li}{\mbox{$^7{\rm Li}\,\,$}}

\newcommand{\simgt}{\lower.5ex\hbox{$\; \buildrel > \over \sim \;$}}
\newcommand{\simlt}{\lower.5ex\hbox{$\; \buildrel < \over \sim \;$}}

\begin{document}
\title{Planet accretion and  the abundances of lithium isotopes}
\author{ J.~Montalb\'an  \inst{1,}\inst{2}, R.~Rebolo\inst{1,}\inst{3} }
\offprints{J.~Montalb\'an}
\institute{Instituto de Astrof\'{\i}sica de Canarias, E-38200-La Laguna (Tenerife), Spain \\
\email{(jmontalb, rrl)@ll.iac.es}
\and Osservatorio Astronomico di Roma, I--00040 Monteporzio, Italy\\
\email{montalbn@coma.mporzio.astro.it}
\and Consejo Superior de Investigaciones Cient\'{\i}ficas. Spain }

   \date{Received December 17, 2001; accepted March 5, 2002}

\abstract{
Planet accretion onto solar type stars may significantly change the stellar 
atmospheric abundances of  $^6$Li and $^7$Li if it
takes place after the star has arrived at the main 
sequence. Ingestion of planets at an earlier phase will not affect theses 
abundances because of extensive pre-main sequence mixing and burning. We present
  quantitative estimates of the main sequence evolution of stellar surface lithium
   abundances after planet ingestion.
At solar metallicities, for stellar masses in the range  
$\sim 1.3-1.1$~${\rm M}_{\odot}$, a large
fraction of the ingested   $^6$Li is
likely to be preserved during the whole main sequence lifetime according to the 
standard model. Preservation of the less fragile $^7$Li isotope occurs in a larger
mass range $\sim 1.3-0.9$~${\rm M}_{\odot}$. At high metallicities typical of planet
 host stars, the ranges of masses are slightly
reduced essentially due to the thicker convective zones. Further reduction is 
expected if non-standard processes cause extra-mixing of material below the base
 of the convective zone, but even in this case there are stellar masses where both
  isotopes are significantly preserved. We conclude
that there is a range of stellar  effective temperature where migration and 
accretion of planets and planetary material  can be empirically tested using
 high-resolution spectroscopy of the lithium isotopes.
}

\maketitle

\section{Introduction}
\label{intro}
The accretion of planets or planetary material onto stars
have been considered  as a possible explanation of the high metallicity frequently 
 found in planet host stars  (Laughlin \& Adams 1997;  Gonzalez 1997, 1998, 2001; 
 Sandquist et al. 1998; Laughlin 2000; Santos et al. 2000; Murray et al. 2001; 
 Pinsonneault et al. 2001).
The very recent detection of the \lil isotope  in the atmosphere of the solar-type 
star  HD~82943 (Israelian et al. 2001), which hosts a planetary system 
(Mayor et al.~2001),  has also been interpreted as evidence for accretion of planets. 
Jupiter-like planets are expected to preserve their original content of the  lithium 
isotopes, while stars like HD~82943 with a mass  $\sim 1.1$~\msol  destroy, 
via nuclear reactions with protons, all its \lil (and a small fraction of the
 more robust $^7$Li)  during the pre-main sequence (PMS) evolution 
 (see e.g. Proffitt and Michaud 1989). It has been argued that ingestion 
 and dilution of planetary material in HD 82943 after its  superficial 
 convective zone  receded to the main-sequence location allowed   long-term 
 preservation of the newly acquired abundances of  lithium isotopes. In 
 fact, according to standard models for this type of stars, the base of the 
 convective zone does not reach the temperatures needed  for  \lil and \li 
 burning ($T \geq 2.2 \times 10^6$~K and $T \geq 2.6 \times 10^6$~K, respectively).
The rate of destruction of \lil by protons has a temperature dependence 
similar to that of $^7$Li, but on much shorter time scale (i.e. 100 times 
shorter at 1~\msol in the standard model). 

In this letter, we examine in quantitative terms the possibility that the 
dilution of one or several Jupiter-like planets in the convective envelope 
of a main sequence star could appreciably change the observed abundance of  
\lil and \li. In the framework of the standard model we predict the   evolution 
of these isotopic  abundances in stars with masses ranging from 
 1.3 to 0.6~${\rm M}_{\odot}$. We also discuss the  effect due to some transport 
 processes which cause extra-mixing below the convective zone (see for review
  Michaud \& Charbonneau 1991). 
We will show that there is a narrow range of mass (or $T_{\rm eff}$)  where the accretion of  planetary material could produce an observable increase of $^{6,7}$Li abundance during a significant fraction of the lifetime of the star.

\section{Models}

The stellar models have been computed with the updated version of the ATON2.0  
code (Ventura et al. 1998). 
We constructed models  with masses ranging from 0.6 to 1.3~\msol  with a step of 
 $\Delta M=0.05$~${\rm M}_{\odot}$, and followed the evolution from the  pre-main 
 sequence  to 7~Gyr. These models use a mixing length theory (MLT) description of 
 convective transport  with  a parameter of mixing length $\alpha=1.6$, as given 
 by the solar calibration (Ventura et al.~1998; Montesinos et al. 2001).  Two 
different chemical compositions were adopted: a) initial solar helium abundance 
of $Y_{\odot}=0.28$ and heavy-elements mass fraction of  $Z_{\odot}=0.02$ 
([Fe/H]=0.0) (Montesinos et al. 2001), b)~$Z=0.04$, with $\Delta Y/ \Delta Z = 2$, 
that is, [Fe/H]=+0.34.

The code used in our model  computations allows to include  gravitational 
settling of helium  and  thermal
diffusion, but it does not consider the diffusion of metals. Diffusion generally  produces
a reduction of the convection zone, however, Turcotte et al. (1998) showed that the convective envelopes 
in models with metal diffusion are deeper than in models including only helium settling.
Furthermore, their results indicate (see their Fig.4) that for the higher masses considered in this paper 
($M\leq 1.3 \,{\rm M}_{\odot}$)  the mass of the convective envelope 
of a non-diffusion model is actually closer to the full-diffusion one than to a 
model incorporating  He settling only (in fact, the non-diffusion and full-diffusion curves overlap). Because of that, we decided to use stellar structures obtained
without including diffusion. Notice also that, a proper treatment of diffusion (Turcotte et al. 1998) 
has no  significant impact on the evolution of lithium abundance in stars with
$M \leq 1.3 \,{\rm M}_{\odot}$.
The nuclear burning rates of \lil and \li in the star's convective envelope 
were computed for each temporal step in evolution using NACRE nuclear reaction 
tables (Angulo et al. 1999).

 Masses for the engulfed sub-stellar companions ranged, from 
1~${\rm M}_{\rm J}$\footnote {Jupiter mass, M$_{\rm J}=9.55\times 10^{-4} {\rm M}_{\odot}$} to 40~${\rm M}_{\rm J}$. This  upper mass limit was adopted to ensure the complete preservation of \lil (Pozio 1991, Nelson et al. 1993), but, in practice, the maximum mass of the accreted planet was  constrained to be less than  the mass in the convective zone of the star under consideration.

The hypothesis and approximations adopted in our study are: 
{\it i)} planet accretion takes place just after the star arrived on the 
main sequence. The analysis presented here does not change if the planet is
engulfed at any other moment during the main sequence lifetime. Such case can be 
dealt by taking into account that the temperature, mass of the convective envelope 
and the stellar surface lithium abundance will be different since they 
decrease as age increases.
{\it ii)} The planet is  completely dissolved in the stellar 
convective envelope. As shown by numerical simulations (Sandquist et al. 1998),
 the reliability of this assumption depends on the mass of the planet 
 ($M_{\rm Pl}$)  and on  the mass of the convective envelope $M_{\rm CZ}$,  
 and also strongly depends on the details of the internal structure of the 
 planet. The smaller the $M_{\rm CZ}$, the lower the fraction of planet 
 dissolved in the convective zone. That means  that, at least for the higher 
 planet masses under consideration, the quantity of lithium provided by the 
 planet must be reduced in a factor equal to the fraction of the planet 
 consumed in the star's convective envelope.  {\it iii)}  The engulfment 
 of a planet has no significant effect in the structure of the star.  
 {\it iv)} The planet is instantly mixed in the convective zone of the star. 
 Given that the diffusion coefficient   in the convective zone is of the 
 order of $10^{13}{\rm cm}^2{\rm s}^{-1}$, the mixing time is lower than 
 the temporal  step in the evolution computation. 

The stellar models were  followed from the PMS, but 
we will not use the surface abundance of \lil and  \li  given by the models
because it is well known that  they overestimate PMS lithium destruction 
(D'Antona \& Mazzitelli 1997).
 For masses lower than  1.4~\msol it is expected that no \lil exists after
 the PMS (Proffitt \& Michaud 1989). As regards surface \li abundance  
 we adopted the values  observed  in stars of the Pleiades Cluster 
 (Soderblom et al.~1993). So the stellar surface abundances of lithium 
 isotopes at the moment of the engulfment of the planet are:
  $N(^6{\rm Li})/N(H)=0$ and $N(^7{\rm Li})/N(H)=F(T_{\rm eff})_{\rm Pleiades}$. 
These abundances  in the planets are assumed equal to the meteoritic  values. 
That is: $N(^7{\rm Li})_0/N(H)=2.10^{-9}$,
 and $ N(^7{\rm Li})_0/N(^6{\rm Li})_0=12.$

  Following Burrows \& Sharp (1999) we assume an Anders \& Grevesse (1989) 
 solar chemical composition for both the planet and the star. 
 In this case,  the difference between the abundance of the isotope 
  $J$ in the star's convective zone before (s) and after (s+p) the dilution 
  of the planet  can be written  as:

\begin{equation}
\epsilon (J)_{\rm s+p} - \epsilon(J)_{\rm s} = \log \left(\frac {x}{x+1}\cdot(1-f)\right) - \log f\,.
\label{aumentoabun}
\end{equation}
\noindent
Where $\epsilon (J)=\log(N(J)/N(H)) +12$, $x=M_{\rm Pl}/M_{\rm CZ}$, and $f$ is
 the  depletion factor of  $J$ in the convective zone just before the absorption
  of the planet.
 So the change in the stellar surface abundance depends strongly on the factor $f$, and it is evident  from Eq.~(\ref{aumentoabun}) that the lithium abundance  enhancement will be  significant  only if the mass fraction of the isotope $J$
 in the planet is much higher than  in the stellar convective envelope.
If the engulfed planet has lost a significant part of its initial
 content of Hydrogen, the mass fraction of lithium in the planet is larger than 
 in the solar chemical composition case. Therefore, also  
 the enhancement of atmospheric lithium abundance will be larger,
 and the results presented here would be a lower limit.

\begin{figure}
\centering
 \includegraphics[width=8.5cm]{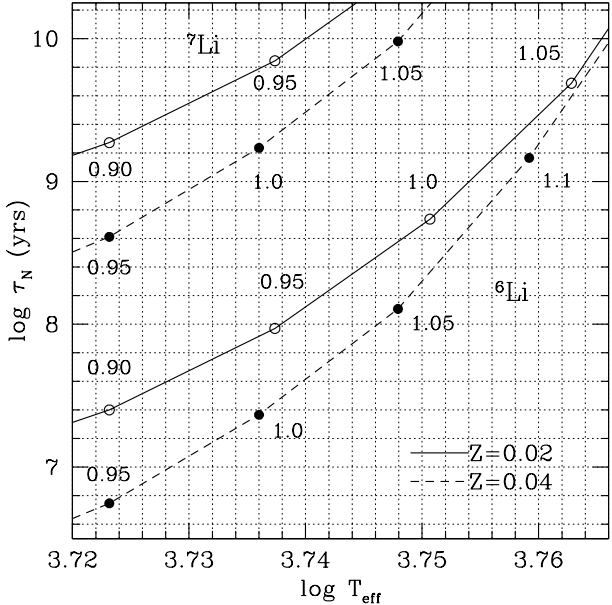}
\caption{\lil (lower curves) and \li (upper curves) lifetime against  
$(p,\alpha)$ nuclear reactions ($N(Li)/N(Li)_0=\exp(-t/\tau_{\rm N})$) as a 
function of effective temperature. The curves correspond to standard models 
at $10^8$~yr with two different metallicities: [Fe/H]=0.0 (solid lines) and 
[Fe/H]=+0.34 (dashed lines). The numbers close to the symbols (empty circles 
for solar metallicity and full circles for [Fe/H]=+0.34) indicate the mass of
 the stellar model.}
\label{tauN}
\end{figure}

 \section{Evolution of Li provided by planetary material}
 
\subsection{Standard model framework}
 
 The standard model assumes that no exchange of material occurs
 between the convective envelope and the stable radiative interior of the star.
  Therefore, the surface chemical composition reflects only the composition of 
  the well mixed convective envelope. In this subsection we present the evolution of 
  lithium in the stellar surface due to the nuclear burning taking place in the 
  convective envelope.
  
 \begin{figure*}
\centering
\includegraphics[width=12cm]{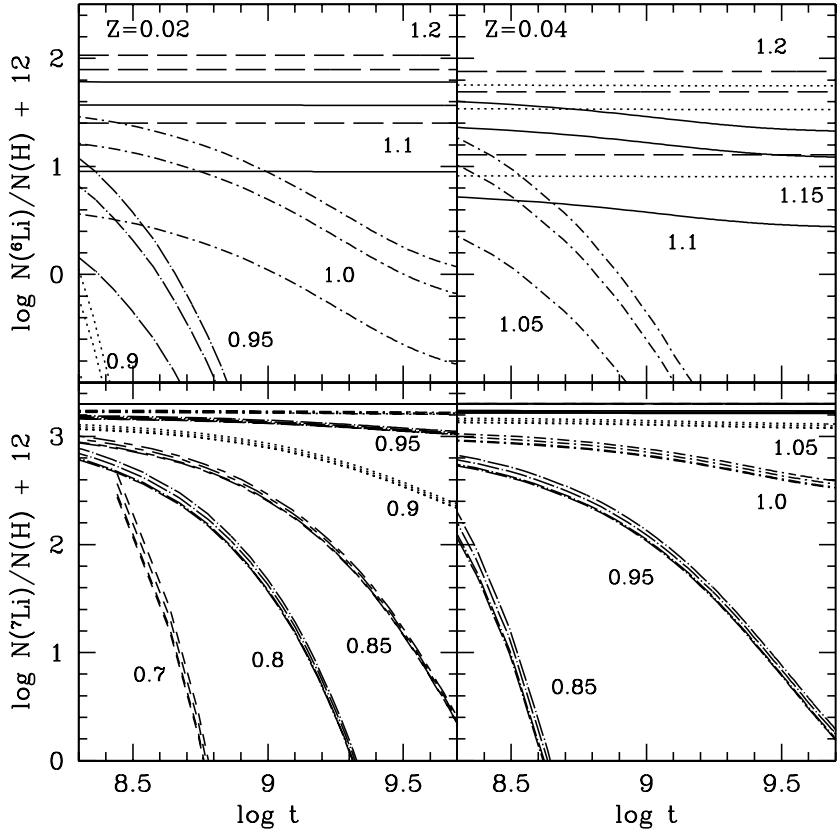}
\caption{Evolution of $^6$Li (upper panels) and $^7$Li (lower panels) provided by 
dissolution of planets in the convective envelope of stars with masses from 1.2 to 
0.7 \msol (numbers indicate the mass of the star) and metallicities :
[Fe/H]=0.0 (left panels) and [Fe/H]=+0.34 (right panels). \lil: for each 
stellar model there are three curves corresponding to (from down to top) 
$\sim$ 1, 5 and 10~M$_{\rm J}$ of planetary material dissolved. \li: as for
 $^6$Li, but we also plot the curve corresponding to no planet accretion. 
 The \li  curves corresponding to : 1.0, 1.1 and 1.2 \msol (left panel), 
 and 1.1, 1.15 and 1.2~\msol(right panel) are concentrated in the top of 
 the respective figures.
}
\label{li67}
\end{figure*}

 Assuming solar chemical composition, a planet of 1~M$_{\rm J}$ supplies 
 $\sim 1.4\,10^{44}$ atoms of $^6$Li \, {\rm and}\,\,$\sim 1.7\,10^{45}$ atoms of $^7$Li. 
 What is the evolution of these atoms during the main sequence lifetime of the star?
The nuclear time scale, $\tau_{\rm N}$, is an estimation of how long these 
atoms could survive to the nuclear burning in the convective envelope. $\tau_{\rm N}$, 
is defined as  the time needed to decrease the superficial abundance of the
 the isotope $J$ by a factor $e$, and it has been obtained from the nuclear 
 reaction rates at each layer in the convective zone, and assuming instantaneous mixing.
 In Fig.~\ref{tauN} we plot $\tau_{\rm N}$ for \lil and \li at 100~Myr 
 as a function of effective temperature.
  We plot  the curves corresponding to two different values of metallicity. We see that for $\log \teff \ge 3.76$  ($\sim$~5700~K), a large fraction of the  \lil from the planet will survive for a long time   in the convective envelope of the star. The value of $\tau_{\rm N}$ given in the plot is a minimum, since 
 during  the main sequence lifetime, the temperature at the bottom of the convective zone decreases, and so the nuclear   time scale increases.  
The effect of increasing the metallicity is to reduce the nuclear  time scale at a given $T_{\rm eff}$. A factor 2 higher metallicity implies a factor 3 lower  time scale at 5500~K, a factor 2 at 5700~K and only a reduction of 10\% 
 for \teff higher than 5800~K.

 Fig.~\ref{li67}   presents the  evolution  (based on models and hypothesis described in
 Sect.~2) of \lil and \li  abundance  for several  stellar masses, 
 and for two values of metallicity: [Fe/H]=0.0 (left panel) and [Fe/H]=+0.34 
 (right panel).
 That temporal dependence is obtained  computing the nuclear 
 depletion of \lil and \li in the stellar convective
 envelope for each temporal step in the evolution and assuming instantaneous mixing.
 We plot only values of lithium abundances corresponding to times
 longer than the time of planet accretion, and lower than the drag-up of the
 convective envelope.
 In the upper panels, we plot the curves corresponding to the \lil that 
 result from the accretion of a planet with $\sim$ 1, 5, and 10~M$_{\rm J}$ 
 in stars with masses between 0.9 and 1.2~M$_{\odot}$    (for 1.25 and 
 1.3~M$_{\odot}$ the enhancement 
  of \lil is slightly larger because the mass of the convective envelope is smaller,
  and  the curves are parallel to those of 1.2~M$_{\odot}$).
 This preliminary and simple modelling provides strong support for
   the interpretation  by Israelian et al. (2001) of the \lil detection in HD~82943. According to our models, assuming   a \teff $\sim 6000$~K, age of  5~Gyr and metallicity [Fe/H]$\sim 0.32$ (Santos et al. 2000) this star has a  mass of $\sim 1.2$~${\rm M}_{\odot}$.
As can be seen in Fig~\ref{li67} right panel, it should preserve any  ingested \lil during all the main sequence lifetime.

The \lil abundances plotted in Fig.~\ref{li67} were obtained assuming   
a complete dissolution of
the planet in the stellar  convective zone. For solar metallicity, 
the masses of the convective envelope in stars with  1.2, 1.1 and 1.0~\msol are  
0.006~M$_{\odot}$, 0.017~M$_{\odot}$ and 
0.034~M$_{\odot}$, respectively. Numerical simulations by Sandquist et al.~(1998) 
indicate that 
as the mass of the convective envelope decreases, the fraction of planet
that is dissolved is reduced. If only 0.3 of a Jupiter-like planet is dissolved 
in the convective zone of a 1.2~\msol star, the corresponding curve of \lil  
will only decrease in $\sim 0.2$ dex. 

Fig.~\ref{li67}  shows that there is a  mass range in which \lil could survive. In fact, for stars with  masses larger than 1.1~\msol (for $Z_{\odot}$, or 1.15~\msol for $Z=2\,Z_{\odot}$) the superficial abundance of \lil does not change with
age. The reason is that   during their main sequence lifetime, the  temperature at the base of the convective zone is not high enough to produce  efficient burning of $^6$Li. As the stellar mass decreases, the depth of the convective zone increases, and then, even if a large quantity of planetary material is dissolved in the stellar envelope, \lil will be depleted in a short time scale ($\sim 10^8$~yr).

The lower panels of Fig.~\ref{li67} presents the predictions for $^7$Li according to standard model. Note that, opposite to $^6$Li, the \li  isotope  is  depleted on the PMS only by a small factor,  and that stars with masses larger than 0.9\msol will not deplete  a significant amount during the main sequence lifetime.
 If these stars were able to dissolve 10~M$_{\rm J}$ in their convective envelope, the effect on the abundance would be to introduce only a small enhancement,  0.2~dex at most.  
The reason is shown in Eq.~(\ref{aumentoabun}), where we see that the increase of abundance depends strongly on the depletion factor of \li in the convective envelope before the accretion of the planetary material.

\subsection{Impact of non-standard transport processes}

 Observations of lithium  in  open clusters  show that the  atmospheric abundance decreases during the main sequence lifetime.
 In order to explain this main-sequence  depletion different mixing processes   are invoked (e.g. Michaud \& Charbonneau 1991), and 
we should consider their effect on  the  evolution of \lil in the stellar mass range where  standard models predict complete preservation. Among these 
processes are: a) microscopic diffusion, that consists roughly in the sink of heavy elements with respect to the light ones due to the gravitational potential; b) turbulent diffusion  which causes mixing of  material between the convective zone and the deep layers where the temperature is high enough to burn lithium.

 Concerning the first one, the diffusion velocities are similar for \lil and $^7$Li, 
 but effects of microscopic diffusion are negligible for the effective temperature 
 and  ages  considered here  (e.g. Richer \& Michaud 1993).  Turcotte et al.~(1998)
 showed that for 1.3~\msol star, the effect of microscopic diffusion on Li is 
 at most a depletion by a factor 0.55, and only after 2 Gyr.
In order to test the effect of turbulent mixing  we have considered a 
diffusion coefficient produced by internal waves  that fits  the evolution of \li 
abundances between the Pleiades and M67 clusters  quite well 
(Montalb\'an \& Schatzman 2000):
 \lil would decrease only in 0.2~dex after 5~Gyr in a 1.2~\msol star ($Z_{\odot}$). 
 However, in a 1.1~\msol ($Z_{\odot}$) star,  \lil will decrease 0.6~dex in the first 
 Gyr and  1.7~dex after 5~Gyr.
Of course, these results depend strongly on the behaviour of the diffusion
 coefficient close to the base of the convective zone. We recall that the diffusion 
 coefficient produced by internal waves has large amplitude at the CZ boundary,
  and then it decreases rapidly downwards. Therefore, the results are very sensitive
   to the distance between the base of the convective zone and the \lil burning layer.
     We have also tested a parametric turbulent diffusion coefficient 
	 $D_{\rm T}=D_0(\rho_{cz}/\rho)^3$ as proposed by Proffitt \& Michaud (1991) 
	 ($\rho_{cz}$ is the density at the base of convective zone). A value of 
	 $D_0 \sim 1.5\,10^3\,{\rm cm}^{2}{\rm s}^{-1}$  produces, after 5~Gyr, a
	  \li depletion of : 1.6~dex for 1~\msol ($\sim 5800$~K), 0.2~dex for 1.1~\msol 
	  ($\sim 6000$~K), and $< 0.1$~dex for 1.2~\msol ($\sim 6100$~K) 
	  (these values are in good agreement with estimations by Pinsonneault et al. 2001
	   using mixing induced by rotation), and  causes a \lil depletion of 0.9~dex at 
	    \teff $\sim$6000~K, 0.05~dex at \teff $\sim$6100~K, and a complete depletion
		 after 1~Gyr for  \teff $\sim$5800~K.

From observations it is inferred that the efficiency of non-standard mixing decreases with the age in the main-sequence. So  
 what happens if the accretion of the planetary material takes place in a \li depleted star at a sufficently old age for extra-mixing to be inefficient? The stellar \li abundance could be significanly enhanced for a long period of time.
For instance, if  a 1~$M_{\rm J}$ planet were dissolved in our Sun's convective envelope would produce an enhancement of    0.3~dex, if  10~$M_{\rm J}$ were accreted the corresponding increase would be 0.9~dex.

\section{Conclusions}
The results presented here show that the accretion of  planetary material  
by a solar-type star will produce an enhancement of the stellar surface abundance 
of lithium which could be observable during a large fraction of the main sequence 
lifetime.  Within the framework of the standard model, the \teff range where the 
ingested \lil could be detected is 5900-6400~K.   The upper limit is imposed by 
 observational evidence given by the Li-Gap in the Hyades (Boesgaard \& Tripicco 1986).
 The lower  limit is the 
result of direct convective mixing and should increase about 100 or 200~K when 
extra-mixing processes are considered. The \teff range where \li would survive 
is larger.
Therefore, a very valuable test for planetary migration/accretion scenarios can be
 provided by 
high-resolution spectroscopic observations of $^{6,7}$Li in planet host stars.

\end{document}